\documentclass[pre,twocolumn,noshowpacs,superscriptaddress]{revtex4-1}

\usepackage{graphicx}
\usepackage{tabularx}

\newcommand{\eqref}[1]{(\ref{#1})}

\newcommand{\Eqref}[1]{Eq.~\eqref{#1}}
\newcommand{\Appref}[1]{Apendix~\ref{#1}}
\newcommand{\Figref}[1]{Fig.~\ref{#1}}
\newcommand{\Tabref}[1]{Table~\ref{#1}}

\newcommand{\latin}[1]{{\itshape #1}}

\newcommand{\denovo}{\latin{de novo}}
\newcommand{\eg}{\latin{e.\,g.}}
\newcommand{\etal}{\latin{et al.\/}}

\newcommand{\ie}{\latin{i.$\,$e.}}

\newcommand{\viz}{\latin{viz.}}

\newcommand{\code}[1]{{\sc #1}}
\newcommand{\DLMESO}{\code{dl\_meso}}
\newcommand{\PACKMOL}{\code{packmol}}

\newcommand{\ang}[1]{{#1^\circ}}

\newcommand{\rc}{r_c}
\newcommand{\Gtransfer}{G_{\mathrm{transfer}}}
\newcommand{\logP}{\log P}
\newcommand{\Nmap}{N_m}
\newcommand{\vmol}{v_m}

\newcommand{\kt}{\kappa_T}
\newcommand{\kB}{k_{\mathrm{B}}}
\newcommand{\kT}{\kB T}

\newcommand{\kb}{k_b}
\newcommand{\re}{r_0}
\newcommand{\ka}{k_a}

\newcommand{\angstrom}{\text{\AA}}
\newcommand{\angscube}{\angstrom{}^{\!3}}

\newcommand{\CnEm}{C$_n$E$_m$}

\newcommand{\NEG}{$-$}
\newcommand{\POS}{\phantom{$-$}}
\newcommand{\px}{\phantom{.0}}
\newcommand{\x}{\phantom{0}}
\newcommand{\xx}{\phantom{00}}

\begin{document}

\title{Dissipative particle dynamics: systematic parametrization using
  water-octanol partition coefficients}

\author{Richard L. Anderson}
\email{richard.anderson@stfc.ac.uk}

\author{David J. Bray}

\affiliation{STFC Hartree Centre, Scitech Daresbury, Warrington, WA4
  4AD, UK}

\author{Andrea S. Ferrante}

\affiliation{Novidec Ltd., 3 Brook Hey, Parkgate, Neston, CH64 6TH,
  UK}

\affiliation{Ferrante Scientific Ltd., 5 Croft Lane, Bromborough, CH62
  2BX, UK}

\author{Massimo G. Noro}

\author{Ian P. Stott}

\author{Patrick B. Warren}
\email{patrick.warren@unilever.com}

\affiliation{Unilever R\&D Port Sunlight, Quarry Road East, Bebington,
  CH63 3JW, UK}

\date{June 27, 2017}

\begin{abstract}
We present a systematic, top-down, thermodynamic parametrization
scheme for dissipative particle dynamics (DPD) using water-octanol
partition coefficients, supplemented by water-octanol phase equilibria
and pure liquid phase density data.  We demonstrate the feasibility of
computing the required partition coefficients in DPD using brute-force
simulation, within an adaptive semi-automatic staged optimization
scheme.  We test the methodology by fitting to experimental partition
coefficient data for twenty one small molecules in five classes
comprising alcohols and poly-alcohols, amines, ethers and simple
aromatics, and alkanes (\ie\ hexane).  Finally, we illustrate the
transferability of a subset of the determined parameters by
calculating the critical micelle concentrations of selected alkyl
ethoxylate surfactants, in good agreement with reported experimental
values.
\end{abstract}

\pacs{Valid PACS appear here}


\maketitle

%\end{CJK*}

\section{Introduction}\label{sec:intro}
In this work we describe a systematic, top-down parametrization scheme
for dissipative particle dynamics (DPD) based on matching
water-octanol partition coefficients, water-octanol phase equilibria,
and pure liquid phase density data.  Apart from water, the DPD `beads'
in the model represent molecular fragments, and as such can be
assembled to cover a wide range of organic materials such as polymers,
surfactants, oils, and so on.  The resulting DPD parameter set can be
used for an equally wide range of applications, and we give the
example of calculating the critical micelle concentrations of alkyl
ethoxylate surfactants.  Our approach is extensible in the sense that
it is easy to broaden the molecular `palette', and flexible in the
sense that the calculations can be re-run semi-automatically if it is
necessary to make different changes to the coarse-graining or include
extra moieties or interactions.  The remaining manual aspects would be
potentially amenable to machine learning approaches. Finally, turned
around, our approach offers a potential novel way of calculating
partition coefficients for new molecules.

We organize the paper by first describing the partition coefficients
that we use as the principal parametrization target.  We then explain
the coarse-grained DPD model and parametrization strategy.  We finally
analyze and discuss the results.  Technical details of the
computational approach are given in a `methods' Appendix, and a
further short Appendix discusses compressibility matching in DPD.

\section{Partition coefficients}
The partition coefficient of an uncharged solute molecule is the ratio
of the molar concentrations in a pair of coexisting bulk phases, at
equilibrium.  The bulk phases themselves are typically made up from a
pair of (near) immiscible solvents, and the corresponding partition
coefficient is usually reported as a base 10 logarithm,
\begin{equation}
  \log P_{A/B}=\log_{10} \frac{[S]_A}{[S]_B}\label{eq:logP}
\end{equation}
where $[S]_A$ and $[S]_B$ are the molar concentrations of a solute
molecule in the two phases, A and B.

By far the most commonly studied partition coefficient is for the
water/octanol system (where octanol means 1-octanol in the present
work); hereafter we shall denote this specific partition coefficient
as simply $\logP$. The partition coefficient is a measure of the
propensity of a solute to partition between hydrophobic and
hydrophilic environments, and is widely used across numerous
application areas such as toxicology, pharmaceutical drug delivery
(pharmacokinetics), and so on \cite{logpuses,sangster_1997}.  For
example, hydrophobic molecules with $1\alt\logP\alt5$ are generally
considered to be cytotoxic since they are able to cross hydrophobic
cell membranes whilst retaining sufficient water solubility to be
active \cite{resistance}.

Given the diverse range of applications, it is not surprising that a
large amount of experimental ${\log P}$ data is available both in the
primary literature and in curated databases, for a wide variety of
solute molecules \cite{shakeflask, sangster_1997, chemspider}.
Moreover, diverse numerical methods with varying degrees of
sophistication and accuracy have been developed to calculate $\logP$
values, to augment the existing experimental data.  These methods
include quantitative structure-property relationships (QSPR)
\cite{qpsr1, qpsr2}; and atom-based \cite{alogp_1998}, and quantum
chemistry motivated methods such as COSMOtherm/COSMO-RS~\cite{logp3,
  cosmors1, cosmors2}.

\begin{table*}
\caption{Coarse-grained (CG) representations of molecules considered
  in the present work.  The CG bead content is denoted by the contents
  of square brackets.\label{beading}}
\vspace{1ex}
\renewcommand{\arraystretch}{0.75}
\setlength\tabcolsep{0.5em}
\begin{tabularx}{0.95\textwidth}{lclccl} 
\hline\\[-3pt]
molecule              && {\sc smiles} code & {\#} beads && CG mapping \\[3pt]
\hline\\[-3pt]
%water                  & - & 1 && [2(H$_2$O)]  \\ 
%
hexane                && cccccc &4 && [CH$_3$][CH$_2$CH$_2$]$_2$[CH$_3$]  \\
octane                && cccccccc & 5 && [CH$_3$][CH$_2$CH$_2$]$_3$[CH$_3$]  \\ 
decane                && cccccccccc & 6 && [CH$_3$][CH$_2$CH$_2$]$_4$[CH$_3$]  \\
dodecane              && cccccccccccc & 7 && [CH$_3$][CH$_2$CH$_2$]$_5$[CH$_3$]  \\
tetradecane           && cccccccccccccc & 8 && [CH$_3$][CH$_2$CH$_2$]$_6$[CH$_3$]  \\[3pt]
benzene               && c1ccccc1  & 3 && [aCHCH]1[aCHCH][aCHCH]1  \\
ethanol               && cco & 2 && [CH$_3$][CH$_2$OH]  \\
1-butanol             && cccco & 3 && [CH$_3$][CH$_2$CH$_2$][CH$_2$OH]   \\ 
1-hexanol             && cccccco & 4 && [CH$_3$][CH$_2$CH$_2$]$_2$[CH$_2$OH]   \\
1-octanol             && cccccccco & 5 && [CH$_3$][CH$_2$CH$_2$]$_3$[CH$_2$OH]  \\[3pt]
butan-1,4-diol        && occcco & 3 && [CH$_2$OH][CH$_2$CH$_2$][CH$_2$OH]  \\
glycerol              && c(c(co)o)o & 3 && [CH$_2$OH]$_3$  \\
tetritol              && occ(c(co)o)o & 4 && [CH$_2$OH]$_4$ \\
ethylamine            && ccn & 2 && [CH$_3$][CH$_2$NH$_2$]  \\
butylamine            && ccccn & 3 && [CH$_3$][CH$_2$CH$_2$][CH$_2$NH$_2$]\\[3pt]
ethanolamine          && occn & 2 && [CH$_2$OH][CH$_2$NH$_2$] \\
diethyl ether         && ccocc & 3 && [CH$_3$][CH$_2$OCH$_2$][CH$_3$]  \\
glyme                 && coccoc & 2 && [CH$_2$OCH$_3$] [CH$_2$OCH$_3$] \\
diglyme               && coccoccoc & 3 && [CH$_3$OCH$_2$][CH$_2$OCH$_2$][CH$_3$OCH$_2$] \\
tetraglyme            && coccoccoccoccoc & 5 && [CH$_3$OCH$_2$][CH$_2$OCH$_2$]$_3$[CH$_3$OCH$_2$]  \\[3pt]
2-hexyloxyethanol     && ccccccocco & 5 && [CH$_3$][CH$_2$CH$_2$]$_2$[CH$_2$OCH$_2$][CH$_2$OH]  \\
ethyl diglyme         && ccoccoccocc & 5 && [CH$_3$][CH$_2$OCH$_2$][CH$_2$OCH$_2$][CH$_2$OCH$_2$][CH$_3$]  \\
phenylmethylether     && cocc1ccccc1 & 4 && [CH$_3$OCH$_2$][aCHCH]1[aCHCH][aCHCH]1\\
phenylpropanol        && occcc1ccccc1 & 5 && [CH$_2$OH][CH$_2$CH$_2$][aCHCH]1[aCHCH][aCHCH]1\\
phenylpropylamine && ncccc1ccccc1 & 5 && [CH$_2$NH$_2$][CH$_2$CH$_2$][aCHCH]1[aCHCH][aCHCH]1  \\[3pt]
\CnEm  && $\mathrm{[c]}_n\mathrm{[occ]}_m\mathrm{[o]}$ & $\frac{1}{2}n+m+1$ && [CH$_3$][CH$_2$CH$_2$]$_{{n}/{2}-1}$[CH$_2$OCH$_2$]$_m$[CH$_2$OH] \\[6pt]
% check brackets and $ signs -- done PBW 14:06 23-06-17
%
\hline
\end{tabularx}
\end{table*}
%\endgroup

Computer simulations, \eg\ molecular dynamics (MD) or Monte Carlo (MC)
methods using atom-based potential functions or coarse-grained
force-fields such as the MARTINI force field \cite{MARTINI-REVIEW},
provide an additional route to the prediction of $\logP$ for small or
medium size solute molecules \cite{aacg,
  POTOFF_2012_ALKANES,Potoff-2012,nitroaro, md2011,
  md2016,kremer_2015,Essex-92, garrido_2011, garrido_ALKANES,
  Garrido_2011-TI-MANYMOLS,LogP-EEM,Maginn-2011,Maginn-EEM-conf-sampl,chen_gemc_2000,
  chen_gemc_2006,5_bannan_calabro_kyu_mobley_2016}.  There have also
been attempts to use multiscale simulation methods mixing atomistic
and coarse-grained potentials \cite{Essex-multilevel,aacg}.  To our
knowledge, all published results to date using these methods resort to
the thermodynamically equivalent definition $\logP=-\log_{10}
e\times{\Delta\Gtransfer}/{RT}$ where $\Delta\Gtransfer$ is the Gibbs
free energy to transfer one mole of solute from phase A to phase B,
and $RT$ is the product of the gas constant $R$ and the temperature
$T$.  The transfer free energy can be formally resolved into the
difference between solvation free energies for which thermodynamic
perturbation and thermodynamic integration methods are typically
employed \cite{Chipot-FE_rev, vanGunsteren-FE-rev,chipotbook,
  Essex-92, garrido_2011, garrido_ALKANES, Garrido_2011-TI-MANYMOLS,
  kremer_2015, fraaije_2016,
  LogP-EEM,Maginn-2011,Maginn-EEM-conf-sampl,
  POTOFF_2012_ALKANES,Potoff-2012,nitroaro}.

It is important to note that a direct application of MC methods such
as Widom insertion are not usually viable for atomistic and coarse
grained simulation methods based on hard core or Lennard-Jones
potentials at liquid densities \cite{widom_1963,frenkelsmit_2001}.
Also determination of the transfer free energy as the
\emph{difference} in solvation free energies may require computing the
latter with a high degree of precision and accuracy.  This makes these
methods very demanding.

In carrying out simulations one could employ pure solvent boxes
(\ie\ pure water and pure octanol), but it is known that the results
for `dry' octanol (\ie\ pure octanol) may be drastically different
from `wet' octanol, since the solubility of water in octanol is
considerable and wet octanol better represents the experimental
situation \cite{chen_gemc_2006}.  More realistically therefore, one
should equilibrate the solvent boxes, and to address this problem one
can turn to Gibbs ensemble methods in which MC solvent molecule
exchange moves are allowed between two separate simulation boxes.
With a Gibbs ensemble method, one can in principle include solute
molecules, and directly evaluate $\logP$ as the ratio of solute
concentrations using the definition in \Eqref{eq:logP}.

Finally we note that the definition of $\logP$ can be extended to
include the molecular components of the solvents themselves.  This is
because we can imagine labelling a small fraction of a given solvent
molecule, and defining its partition coefficient using \Eqref{eq:logP}
where the ratio $[S]_A/[S]_B$ can be replaced by the ratio of the
molar compositions in the coexisting phases.  For example, in the
water-octanol system we can define water and octanol partition
coefficients, and use them interchangeably with the mutual
solubilities.  Obviously this only makes sense for the equilibrated
system (\ie\ `wet' octanol as defined above).

\section{Coarse grained model definition}
In our approach the DPD beads represent molecular fragments comprising
1--3 `heavy atoms' (\ie\ C, O, N in this work), with the exception of
water (H$_2$O) which is treated super-molecularly.  This means that a
wide variety of both aqueous and non-aqueous systems can be modelled
by combining these fragments as a kind of molecular `Lego' game.  It
also means the approach is extensible, since the molecular palette is
easily enlarged.

To establish the basis for the above coarse-graining scheme, we first
follow Groot and Rabone in defining a water mapping number, in our
case $\Nmap=2$ so that each water bead corresponds, on average, to two
water molecules.  Following well established protocols we also assert
that the density of water in our model corresponds to $\rho\rc^3=3$ in
DPD units.  We can then use the mapping number tautology
$\rho\Nmap\vmol\equiv1$, where $\vmol\approx 30\,\angscube$ is the
molecular volume of liquid water (corresponding to a molar volume
$\approx36.0\,\mathrm{cm}^3\,\mathrm{mol}^{-1}$), to determine that $\rc
\approx 5.65\,\angstrom$.  This underpins the conversion of all
lengths and molecular densities in the model.

In our model, alkane molecules are constructed from connected (bonded)
beads comprising (i) $\mathrm{CH_2CH_2}$ groups of atoms and (ii)
$\mathrm{CH_3}$, a terminal methyl group. Similarly alcohol molecules
are constructed by bonding together alkane beads and a specific bead
containing an alcohol functionality, \eg\ comprised of the
$\mathrm{CH_2OH}$ group of atoms. Amine molecules follow the same
model definition where the amine functionality is captured in a bead
comprised of $\mathrm{CH_2NH_2}$ atoms. Benzene rings are constructed
from a total of three beads, each comprising $\mathrm {CHCH}$ groups,
bonded together in a triangle.  We name the corresponding DPD beads
$\mathrm{aCHCH}$, to remind that they are part of an aromatic
ring. Ether beads are designated as being formed from
$\mathrm{CH_2OCH_2}$ or $\mathrm{CH_3OCH_2}$ groups, depending on the
location of the bead (central or terminal). Using two bead types for
the ethers is essential for reproducing the correct ${\log\, P}$ for
diglyme, for example. Atom to beaded structures are given in detail in
\Tabref{beading}.

Having decided on the level of coarse graining, the next part of the
model definition is to specify the bonded and non-bonded interactions
between beads.  For the non-bonded interactions we take the usual DPD
pairwise soft repulsion, $\phi=\frac{1}{2}A_{ij}(1-r/R_{ij})^2$ for
$r\le R_{ij}$ and $\phi=0$ for $r>R_{ij}$.  The values of $A_{ij}$ and
$R_{ij}$ are set by the bead types, and the determination of these
interaction parameter matrices is the central problem addressed in the
present work.  Note that unlike much of the existing literature we
allow the repulsion ranges to differ from the canonical
$R_{ij}\equiv\rc$.  As a baseline though we set $R_{ii}=\rc$ and
$A=25$ for the water bead self-repulsion (see discussion in
\Appref{app:comp}).

For the bonded interactions we take an approach motivated by previous
work and our own experience.  A simple harmonic potential $\phi_b =
\frac{1}{2}\kb(r - \re)^2$ was chosen to represent bonds between
connected DPD beads.  The nominal bond length in this (\ie\ $\re$)
could be adjusted to reproduce the true experimental distances between
groups, and our ambition in future is to do this, however in the
present work we adopt a pragmatic approach in which $\re = 0.1 \times
(N_i + N_j)$ (units of $\rc$) where $N_i$ and $N_j$ are the number of
heavy atoms in beads $i$ and $j$, respectively (see \Tabref{DZ}).
A single bond constant $\kb=150$ was adopted throughout (in units of
$\kT$).  Note that, contrary to the usual practice in MD, we (and
others) do not exclude the non-bonded interaction between two bonded
DPD beads.

Finally, in our model we explicitly introduce an element of rigidity
by including a harmonic angular potential between pairs of bonds.
This appears to be essential for the correctness of molecular models
at the level of coarse graining used here: rigidity has been shown to
be important in a number of DPD studies of small molecules
(surfactants and lipids), for example surfactant tail stiffness
directly controls the surfactant effective length and
area-per-molecule \cite {venturoli_1999-stiffness}, and both
quantities affect surfactant packing and self-assembly
\cite{tail-stiffness-phase}.  We here adopt the same three-body
angular potential used by Smit and collaborators
\cite{venturoli_1999-stiffness,Smit-review},
\viz\ $\phi_a=\frac{1}{2}\ka(\theta-\theta_0)^2$ where $\theta$ is the
angle between the bonds.  In the present work we set $\theta_0 =
\ang{180}$ and $\ka = 5$ (in units of $\kT$) for everything except
benzene rings, where instead $\theta_0 = \ang{60}$ (and $\ka = 5$) is
used. In this current study we consider only linear or ring (in the
case of benzene) molecules.

\section{Staged parametrization scheme}
Having specified the model as above, we now turn to the critical task
of determining a suitable set of interaction parameters.  Here we
adopt a staged approach, in which space of undetermined parameters is
sequentially refined until we have a single consistent set.  This
reduces the problem to a manageable sequence of `unit operations'.

\subsection{Cutoff distances}
In the first step we focus mainly on the cutoff distances $R_{ij}$.
Note that DPD is an example of a `mean-field fluid' and thus there is
an element of trade-off between $A_{ij}$ and $R_{ij}$ since the
properties are expected to be largely determined by $A_{ij}R_{ij}^3$
\cite{ardlewis}.  On this basis one could set $R_{ij}\equiv\rc$ for
all bead types and attempt to accommodate variation in `bead size'
within the $A_{ij}$ matrix.  However under the chosen mapping the
beads contain different numbers of atoms and do contribute slightly
unequally to the total molar volume of the molecules under
consideration.  As already mentioned therefore, we allow ourselves the
flexibility of separately specifying $A_{ij}$ and $R_{ij}$ and use the
self repulsion cutoffs $R_{ii}$ to capture, at `zeroth order' as it
were, the contribution of the molecular fragments to the overall molar
volumes.

To do this we used the rules developed by Durschlag and Zipper (DZ)
for individual atom contributions to molar volume \cite{DZ}.  We then
define $R_{ii}$ such that $R_{ii}^3$ is proportional to the molar
volume of the fragment, with the constant of proportionality being set
by the water bead mapping.  To deal with the cutoff between dissimilar
bead types, we adopt the simple arithmetic `mixing rule'
$R_{ij}=\frac{1}{2}(R_{ii}+R_{jj})$.  (The observant reader will
notice this is equivalent to assigning an effective radius
$R_i=R_{ii}/2$ to individual beads, and asserting that
$R_{ij}=R_i+R_j$.) \Tabref{DZ} presents the molecular fragments used
as DPD beads and their corresponding volumes determined by the DZ
method and \Tabref{tab:interactions} shows the resulting cutoffs.
With this first step the cutoffs are now specified (note that no DPD
simulations have been undertaken thus far).

\begin{table}
\caption{Coarse grained (CG) bead content, number of heavy atoms, and
  molar volumes ($\mathrm{cm}^3\,\mathrm{mol}^{-1}$) calculated from
  Durschlag and Zipper \cite{DZ}. For water, $\Nmap=2$ is the mapping
  number used to define the DPD length scale $\rc$.\label{DZ}}
\setlength\tabcolsep{0.5em}
\vspace{1ex}
\begin{tabularx}{0.48\textwidth}{lcccc}
\hline\\[-6pt]
CG bead && $N_i$ ({\#} heavy atoms) && molar volume\\[3pt]
\hline\\[-6pt]
$\mathrm{[H_2O]_2}$       && 
2\makebox[0pt][l]{ (${}\equiv\Nmap$)} && 36.0   \\
$\mathrm{[CH_3]}$         && 1 && 31.4   \\
$\mathrm{[CH_2CH_2]}$     && 2 && 44.6   \\
$\mathrm{[CH_2OH]}$       && 2 && 33.9   \\
$\mathrm{[aCHCH]}$        && 2 && 32.8   \\
$\mathrm{[CH_2NH_2]}$     && 2 && 38.7   \\
$\mathrm{[CH_2OCH_2]}$    && 3 && 50.1   \\
$\mathrm{[CH_3OCH_2]}$    && 3 && 53.2   \\[3pt]
\hline
\end{tabularx}
\end{table}

\subsection{Self-repulsion}
We now turn to the repulsion amplitude matrix $A_{ij}$.  Again taking
a pragmatic approach we deal with the self-interaction parameters
first, before turning to the repulsion amplitudes between dissimilar
bead types.  In the second parametrization step therefore we make the
initial assumption that $A_{ij}=\frac{1}{2}(A_{ii}+A_{jj})$ and adjust
the self-repulsion $A_{ii}$ to fit the experimental densities (at
atmospheric pressure and 25$^{\circ}$C) of a number of simple
molecular liquids containing these beads, computed using the
methodology described in \Appref{app:meth}.  With this second
step the self-repulsion amplitudes are now fixed.

\subsection{Off-diagonal repulsions}
Finally, in the third parametrization step we turn to the repulsion
amplitudes between dissimilar bead types (the off-diagonal $A_{ij}$
matrix entries).  Here for the first time we target the experimental
$\logP$ values, using the computational methodology described in
\Appref{app:meth}.  We divide the target molecules into a
\emph{training set} of size eight, with a further thirteen used for
testing the model. As optimization targets we used a combination of
water-octanol mutual solubilities and the training set $\logP$ values.

\Tabref{tab:logp} lists the molecules considered in our $\logP$
study and highlights test versus training solute molecules. The
parameters represent the closest fit achieved in terms of the minimum
value for root-mean-square error (RMSE) in a manual fitting procedure
with a target of 0.3 log units for the training set. Note that the
mutual solubilities of water with octanol were both included in the
training set. In order to obtain our training set parameters we first
optimized $A_{ij}$ for the alcohol molecules to achieve the minimum
RMSE for these molecules (note that the self-interactions were held
constant at the values determined in the previous,
liquid-phase-density-matching step). Following this, whilst holding
the alcohol parameters fixed, the amine beads parameters were
optimized. The ethers and benzene were dealt with similarly.
\Tabref{tab:interactions} gives the final set of $A_{ij}$ values
resulting from this procedure.  In combination with the bonded
interaction parameters, this fully specifies the DPD model.

\begin{figure}
\begin{centering}
\includegraphics[width=3.2in]{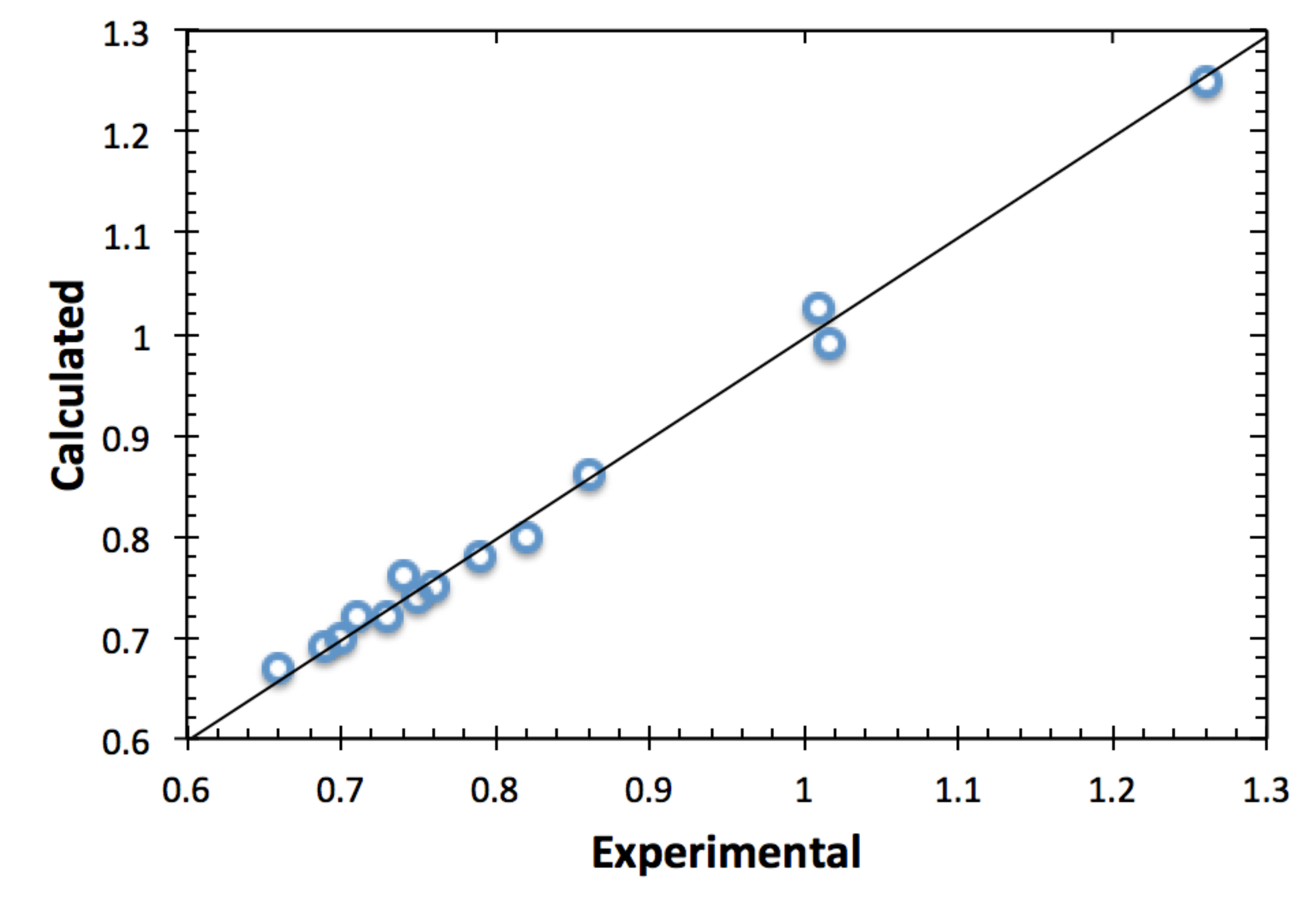}
\end{centering}
\caption{Experimental versus calculated liquid phase densities. Errors
  in each calculated result are smaller than the symbols. Units are
  $\mathrm{g}$ $\mathrm{ cm^{-3}}$. \label{fig:density}}
\end{figure}

\begin{figure} 
\begin{centering}
\includegraphics[width=3.2in]{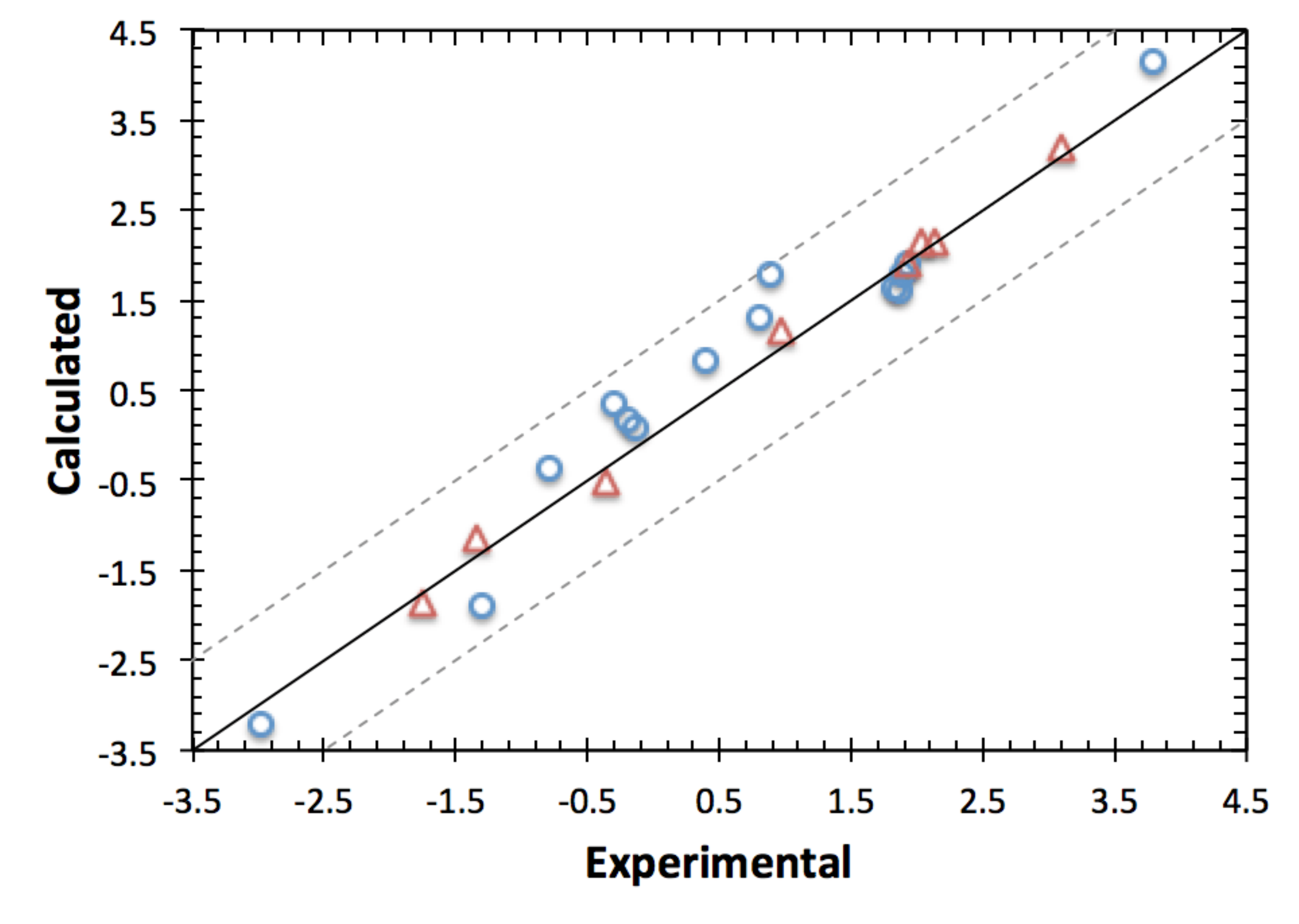}
\end{centering}
\caption{Experimental versus calculated $\logP$ values.  The model has
  a root-mean-square error (RMSE) of 0.20 log units for the training
  set (red triangles) and 0.45 log units for the test set (blue
  circles).  Dashed lines represent $\pm1$ log units.\label{fig:logp}}
\end{figure}

\section{Results}
Figure~\ref{fig:density} compares the calculated densities to the
experimental densities for the species used to parametrize
self-interaction parameters. The foundation for the fitting of the
self-interactions was chosen to be the alkane bead types
($\mathrm{CH_{2}CH_{2}}$ and $\mathrm{CH_{3}}$) as these are present
in most solute molecules considered in our work. Hence, multiple
alkane molecules were sampled in the parametrization of the
self-interaction parameters.  Our results for all molecules are in
excellent agreement with literature values. The deviations for the
alkanes are within 1.5\% of the reported values. The trend reproduced
by our model slightly over estimates the shorter alkanes densities and
slightly underestimates at the longer alkane chain lengths.
\Tabref{density} lists the calculated density versus experimental values.

\Tabref{tab:logp} compares the experimental and calculated $\logP$
values for the solute molecules considered in this study. Overall the
calculated $\logP$ values are in good agreement with experimentally
determined values, all being within 0.6 log units with the exception
of diethyl ether from the test set. Considering the coarseness of the
model adopted (manual optimization of parameters, simple model for
specifying bond lengths, single angular rigidity), is perhaps a
better-than-expected result.  The agreement between calculated and
experimental $\logP$ values can be observed clearly in
\Figref{fig:logp}. Our model produces a RMSE for of 0.20 and 0.45 log
units for our training and test set of molecules, respectively. This
is excellent when compared with other predictions of ensembles of
solutes (albeit with a small data set in this case).

In setting out an appropriate model for reproducing $\logP$ the most
important results to get correct are those of octanol and water
themselves (\ie\ the mutual solubilities). These are therefore very
important molecules in our training set.  All other calculated results
depend upon these values. For example, if there is too much water in
the octanol bulk phase then that whole bulk region is too
hydrophilic. As a consequence, $\logP$ for other solutes will be
skewed. The interaction parameters listed in 
\Tabref{tab:interactions} gives values of 3.2 and $-$1.1 for octanol and
water (logarithm of solubility and inverse solubility,
respectively). Experimental values are 3.1 and $-$1.3 derived from the
mutual solubilities reported in Dallos and Liszi
\cite{dallos1995liquid+}. Our results here are the best that were
possible with our adopted coarse grained model. Attempts to improve
the value for water resulted in poorer values for octanol.

\begin{table}
\caption{Repulsion amplitudes ($A_{ij}$) and cut-off distances
  ($R_{ij}$) between all bead pairs in the model (note the letter `a' in
  `aCHCH' indicates this bead is from an aromatic ring).\label{tab:interactions}}
\renewcommand\arraystretch{0.75}
\setlength\tabcolsep{1em}
\vspace{1ex}
\begin{tabularx}{0.48\textwidth}{llccc}
\hline\\[-3pt]
bead $i$ & bead $j$ && $A_{ij}$ & $R_{ij}$ \\[3pt]
\hline\\[-3pt]
$\mathrm{H_2O}$         &	$\mathrm{H_2O}$	        &&	25.0	&	1.0000	\\
$\mathrm{H_2O}$	        &	$\mathrm{CH_2OH}$	   &&	14.5	&	0.9900	\\
$\mathrm{H_2O}$	        &	$\mathrm{CH_2CH_2}$	   &&	45.0	&	1.0370	\\
$\mathrm{H_2O}$	        &	$\mathrm{CH_3}$	       &&	45.0	&	0.9775	\\
$\mathrm{H_2O}$	        &	aCHCH	               &&	45.0	&	0.9850 	\\[3pt]
$\mathrm{H_2O}$	        &	$\mathrm{CH_2NH_2}$	   &&	14.5	&	1.0120	\\
$\mathrm{H_2O}$         &	$\mathrm{CH_2OCH_2}$	&&	24.0	&	1.0580	\\
$\mathrm{H_2O}$    	    &	$\mathrm{CH_3OCH_2}$	&&	32.0	&	1.0695	\\
$\mathrm{CH_2OH}$	    &	$\mathrm{CH_2OH}$	  &  &	14.0	&	0.9800	\\
$\mathrm{CH_2OH}$	    &	$\mathrm{CH_2CH_2}$	   &&	26.0	&	1.0270	\\[3pt]
$\mathrm{CH_2OH}$	    &	$\mathrm{CH_3}$	        &&	26.0	&	0.9675	\\
$\mathrm{CH_2OH}$	    &	aCHCH            	  &  &	27.0	&	0.9750	\\
$\mathrm{CH_2OH}$	    &	$\mathrm{CH_2NH_2}$	   &&	18.0	&	1.0020	\\
$\mathrm{CH_2OH}$	    &	$\mathrm{CH_2OCH_2}$	&&	25.0	&	1.0480	\\
$\mathrm{CH_2OH}$	    &	$\mathrm{CH_3OCH_2}$	&&	25.0	&	1.0595	\\[3pt]
$\mathrm{CH_2CH_2}$	    &	$\mathrm{CH_2CH_2}$	  &  &	22.0	&	1.0740	\\
$\mathrm{CH_2CH_2}$	    &	$\mathrm{CH_3}$	        &&	23.0	&	1.0145	\\
$\mathrm{CH_2CH_2}$	    &	aCHCH	                &&	27.0	&	1.0220 	\\
$\mathrm{CH_2CH_2}$	    &	$\mathrm{CH_2NH_2}$	  &  &	22.5	&	1.0490	\\
$\mathrm{CH_2CH_2}$	    &	$\mathrm{CH_2OCH_2}$	&&	28.5	&	1.0950	\\[3pt]
$\mathrm{CH_2CH_2}$	    &	$\mathrm{CH_3OCH_2}$	&&	28.5	&	1.1065	\\
$\mathrm{CH_3}$	        &	$\mathrm{CH_3}$	        &&	24.0	&	0.9550	\\
$\mathrm{CH_3}$	        &	aCHCH	                &&	27.0	&	0.9625	\\
$\mathrm{CH_3}$	        &	$\mathrm{CH_2NH_2}$	  &  &	24.0	&	0.9895	\\
$\mathrm{CH_3}$	        &	$\mathrm{CH_2OCH_2}$	&&	28.5	&	1.0355	\\[3pt]
$\mathrm{CH_3}$	        &	$\mathrm{CH_3OCH_2}$	&&	28.5	&	1.0470	\\
aCHCH	                &	aCHCH	                &&	27.0	&	0.9700  \\
aCHCH	                &	$\mathrm{CH_2NH_2}$	  &  &	27.0	&	0.9970  \\
$\mathrm{CH_2NH_2}$	    &	$\mathrm{CH_2NH_2}$	   &&	21.5	&	1.0240	\\
$\mathrm{CH_2OCH_2}$	&	$\mathrm{CH_2OCH_2}$	&&	25.5	&	1.1160	\\[3pt]
$\mathrm{CH_2OCH_2}$	&	$\mathrm{CH_3OCH_2}$	&&	25.5	&	1.1275	\\
$\mathrm{CH_3OCH_2}$	&	$\mathrm{CH_3OCH_2}$	&&	25.5	&	1.1390	\\[3pt]
\hline
\end{tabularx}
\end{table}

\begin{table}
\caption{Experimental versus calculated $\logP$ values for considered
  solute molecules (in $\log_{10}$ units). The category distinguishes
  training set molecules (A) from test set molecules (B).  The error
  (defined as standard deviation in the sample mean for $\logP$) in
  the calculated values is $<10$\% of the calculated mean
  value.\label{tab:logp}}
\renewcommand\arraystretch{0.75}
\vspace{1ex}
\begin{tabularx}{0.48\textwidth}{lcccc} 
\hline\\[-3pt]
%Solute & Train (A) & $\logP^{\text{expt}}$ & $\log\,P^{\text{calc}}$ & $\Delta\logP$ \\
%       & or Test (B)  &              &            &  \\
solute & category & \multicolumn{2}{c}{$\logP$} & $\Delta\logP$ \\
 &  & \POS(expt) & \POS(calc) & \\[3pt]
\hline\\[-3pt]
octanol                & A &  \POS 3.1   & \POS 3.2  & \POS 0.1  \\ 
water                  & A &  \POS 1.3   & \POS 1.1  & \NEG 0.2  \\ 
hexane                 & A &  \POS 3.8   & \POS 4.1  & \POS 0.3  \\  
ethanol                & B &  \NEG 0.3   & \POS 0.3  & \POS 0.6  \\ 
1-butanol              & B &  \POS 0.8   & \POS 1.3  & \POS 0.5  \\[3pt]
1-hexanol              & A &  \POS 2.0   & \POS 2.2  & \POS 0.2  \\
butan-1,4-diol         & B &  \NEG 0.8   & \NEG 0.4  & \POS 0.4  \\
glycerol               & A &  \NEG 1.8   & \NEG 2.0  & \NEG 0.2  \\
tetritol               & B &  \NEG 3.0   & \NEG 3.2  & \NEG 0.2  \\
ethylamine             & B &  \NEG 0.1   & \POS 0.1  & \POS 0.2  \\[3pt]
1-butylamine           & A &  \POS 1.0   & \POS 1.2  & \POS 0.2  \\
benzene                & A &  \POS 2.1   & \POS 2.1  & \POS 0.0  \\
ethanolamine           & B &  \NEG 1.3   & \NEG 1.8  & \NEG 0.5  \\
3-phenyl-1-propanol    & B &  \POS 1.9   & \POS 1.8  & \NEG 0.1  \\
3-phenyl-1-propylamine & B &  \POS 1.8   & \POS 1.6  & \NEG 0.2  \\[3pt]
diethyl ether          & B &  \POS 0.9   & \POS 1.8  & \POS 0.9  \\
glyme                  & B &  \NEG 0.2   & \POS 0.2  & \POS 0.4  \\
diglyme                & A &  \NEG 0.4   & \NEG 0.5  & \NEG 0.1  \\
2-(hexyloxy)ethanol    & B &  \POS 1.9   & \POS 1.6  & \NEG 0.3  \\
ethyl diglyme          & B &  \POS 0.4   & \POS 0.8  & \POS 0.4  \\[3pt]
benzylmethyl ether     & B &  \POS 1.9   & \POS 1.9  & \POS 0.0  \\[3pt]
\hline
\end{tabularx}
\end{table}

\begin{table}
\caption{Experimental versus calculated densities
  ($\mathrm{g}\,\mathrm{cm}^{-3}$) for some pure liquids at atmospheric
  pressure and $25^\circ $C. Statistical errors in the calculated
  values are all in the third decimal place. \label{density}}
\renewcommand\arraystretch{0.75}
\setlength\tabcolsep{1em}
\vspace{1ex}
\begin{tabularx}{0.48\textwidth}{lccc} 
\hline\\[-3pt]
species         & expt & calc & relative deviation \\[3pt]
\hline\\[-3pt]
hexane          & 0.66   & 0.67  &      $+$1.5 \% \\
octane          & 0.70   & 0.70  &      ${}<1$ \% \\ 
decane          & 0.73   & 0.72  &      $-$1.4 \% \\
dodecane        & 0.75   & 0.74  &      $-$1.3 \% \\
tetradecane     & 0.76    & 0.75 &      $-$1.3 \% \\[3pt]
ethanol         & 0.79    & 0.78 &      $-$1.0 \% \\
octanol         & 0.82    & 0.80 &      $-$2.5 \% \\
butan-1,4-diol  & 1.02    & 0.99 &      $-$3.9 \% \\
glycerol        & 1.26    & 1.25 &      $-$0.7 \% \\
diethyl ether   & 0.71    & 0.72 &      $+$1.4 \% \\[3pt]
tetraglyme      & 1.01    & 1.04 &      $+$3.0 \% \\
ethylamine      & 0.69    & 0.69 &      ${}<1$ \% \\
butylamine      & 0.74    & 0.76 &      $+$2.7 \% \\
benzene         & 0.86    & 0.86 &      ${}<1$ \% \\[3pt]
\hline
\end{tabularx}
\end{table}

Our hexane model presented a reasonable match to literature $\logP$
values, and in benzene we achieved a slightly better match. For the
alcohol molecules studied, there is mixed agreement between reported
and calculated values. Generally there is good agreement with the
longer and di-, tri- and tetra-alcohols fitting the best; this is by
construction since in our optimization efforts we focused on the
longer molecules for the alcohol functionality.  Butanol carries a 0.5
log unit difference. The most notable exception is ethanol with a
difference of 0.6 log units, however. Our model presents ethanol as
being slightly favoured in the octanol phase as opposed to the water
phase. This behaviour could be attributed to an unrealistically
favorable interaction between the alcohol functionality bead CH$_2$OH
and the alkane group CH$_2$CH$_2$, or, possibly, to increased
solubility of water in octanol in our model. This could in turn allow
a greater proportion of ethanol to reside in the octanol rich
phase. It would appear a potential solution to this could be to make
the H$_2$O/CH$_2$OH interaction more favorable or the CH$_2$OH/alkane
less favourable. However, both of these approaches has the counter
effect of reducing the quality of the fit for the other, longer
alcohols (and indeed the all important octanol and water values).

We have sampled six solute molecules that contain ether groups. Whilst
our model gives a good fit for diglyme, 2-(hexyloxy)ethanol and ethyl
diglyme, diethyl ether proves to be a poor fit---in fact the poorest
fit of all molecules we have considered.  This trend of poorer
agreement for short molecules and better for long one is again likely
due to the fact that we have chosen to fit the ether functionality to
a longer molecule (diglyme), for better transferability to longer
chains. The fact that our model for diethyl ether is predicted to be
much more hydrophobic than in reality, is due to the lower accuracy of
our coarse grained model for smaller molecules or is related to some
physics we are neglecting in our simplified model for ethoxylated
groups. For diethyl ether our model of a linear molecule may be to
blame. It is possible that a better representation of the average
conformation is given by a bent trimer, possibly with shorter bonds,
which would make it more soluble in water. For the case of glyme, we
end up in a situation similar to that of ethanol, in which the DPD
determined $\logP$ is of the opposite sign (although the difference
from the experimental value is small at 0.4 log units).

In a larger parametrization effort it would be prudent to take into
consideration the conformation of the ethoxylated chains. It is known
that different conformations of chains, with different relative
orientation of the dipole moment centered on the oxygen between
adjacent monomers, result in considerably different solvation free
energies for the molecule. In particular, a recent combined Raman and
density functional theory study \cite{glyme-solub}, has shown that
indeed the trans-gauche-trans (tgt) conformation of the molecule has a
much larger solvation free energy than the trans-trans-trans (ttt)
(which we used to represent the geometry of the coarse-grained dimer),
and that population of the (tgt) conformation in water is 79\%. It is
unclear to us currently whether the simple shortening of the bond
length will be enough to represent better glyme, or if instead a
better transferability is only possible by including more detailed
physics, \eg\ in the form of an internal degree of freedom for the
ether bead or the explicit representation of the ether bead dipole
together with a polar model for the coarse-grained water.

Again a trend of better reproduction of the solubility of longer
molecules versus shorter ones can be seen for the case of the amines.
Butylamine is in good agreement with reported values, and while the
difference between reported and calculated for ethylamine is only
small, the sign is incorrect. Again, a simple solution might perhaps
be to make the amine-water interaction more favourable, but, like in
the alcohol case this has further reaching consequences for other
solutes. For example, ethanolamine whose $\logP$ is already calculated
as too negative by our model. With this latter case we see another
short molecule that shows a large deviation from experimental values.
Three solutes have been explored that combine benzene functionality in
addition to alcohols, amines and ether functionality. All three
molecules have reported $\logP$ values in the region of 1.8--1.9
\cite{shakeflask, chemspider}.  Our model does a good job are
reproducing these values of the three solutes.

Finally, to address the transferability of the model and parameters
developed above, the critical micelle concentrations (CMCs) of seven
non-ionic surfactants of the \CnEm\ family have been calculated as
outlined in \Appref{app:meth}. There is good general agreement
between calculated and experimental values across all surfactant
molecules explored. These results are pleasing indeed and provide a
indication that fitting interaction parameters to $\logP$ values may
allow additional properties to be calculated where relative solubility
of chemical species play an important role, such as self assembly
(micelle formation), phase separation or degree of mixing, and could
provide a good basis for larger scale parametrization efforts of
models with application to surfactant systems.

\begin{table}
\caption{Experimental versus calculated critical micellisation
  concentrations (CMCs) (units of wgt\,\%) for seven surfactants
  of the \CnEm\ family. Experimental values have been extracted from
  several literature sources \cite{cnem1,cnem2,cnem3,cnem4}. Number in
  brackets after calculated result denotes error
  estimate. \label{tab:cmc}}
\setlength\tabcolsep{1em}
\vspace{1ex}
\begin{centering}  
\begin{tabularx}{0.48\textwidth}{lcccc}
\hline\\[-6pt]
surfactant && CMC (expt) && CMC (calc) \\[3pt]
\hline\\[-6pt]
$\mathrm{C_6E_4}$	&& 2\px\xx &&	0.8(3)\xx \\
$\mathrm{C_8E_4}$	&& 0.2\xx &&	0.2(1)\xx \\
$\mathrm{C_8E_8}$	&& 0.5\xx &&	0.3(1)\xx \\
$\mathrm{C_{10}E_6}$	&& 0.04\x &&	0.04(2)\x \\
$\mathrm{C_{10}E_9}$	&& 0.07\x &&	0.05(2)\x \\
$\mathrm{C_{12}E_6}$	&& 0.003 &&	0.008(4) \\
$\mathrm{C_{12}E_7}$	&& 0.003 &&	0.009(5) \\[3pt]
\hline
\end{tabularx}
\end{centering}
\end{table}

\section{Discussion}
In this work we have demonstrated that the DPD method can be used to
for direct (\ie\ brute-force) calculation of $\logP$ values for small
molecules, and thereby used to parametrize the underlying coarse
grained model.  We have also shown how to optimize self-interaction
parameters and cutoff distances based on bead volumes to give good
agreement with experimental liquid phase densities. We stress that we
do not expect the interaction parameters to remain valid if the choice
of bond length is altered, however.  In carrying out the work it has
become apparent how to make optimal use of simulation setup for rapid
equilibration and to achieve reliable results.  The good agreement we
have observed between experimental and calculated partition
coefficients, and critical micelle concentrations for alkyl ethoxylate
surfactants, give us confidence in the resulting parameter set and its
transferability.  An application to the whole phase diagram of the
non-ionic surfactants will be the subject of a forthcoming
communication.

Recent work by, for example Lee \etal, has focused on parametrizing
DPD models by limiting activity coefficients \cite{lee-markov}.  This
should in principle give the same results as the present brute-force
method, under the assumption that the force field recovers the same
water-octanol phase coexistence compositions. However the equivalence
of the two approached depends crucially upon reproducing exactly the
experimental solubility of water in octanol. Even with our very good
result ($-1.1$ vs $-1.3$ for the logarithm of the mutual
solubility of water in octanol), this amounts to $\approx 30$\%
error. Hence the ratio of the mutual limiting activity coefficients of
water and octanol bears the same level of error. In practice it
appears that for current DPD potential models the two approaches are
not exactly equivalent and one has to make a choice on which method to
use for parametrization.

\begin{figure}
\begin{centering}
\includegraphics[width=3.2in]{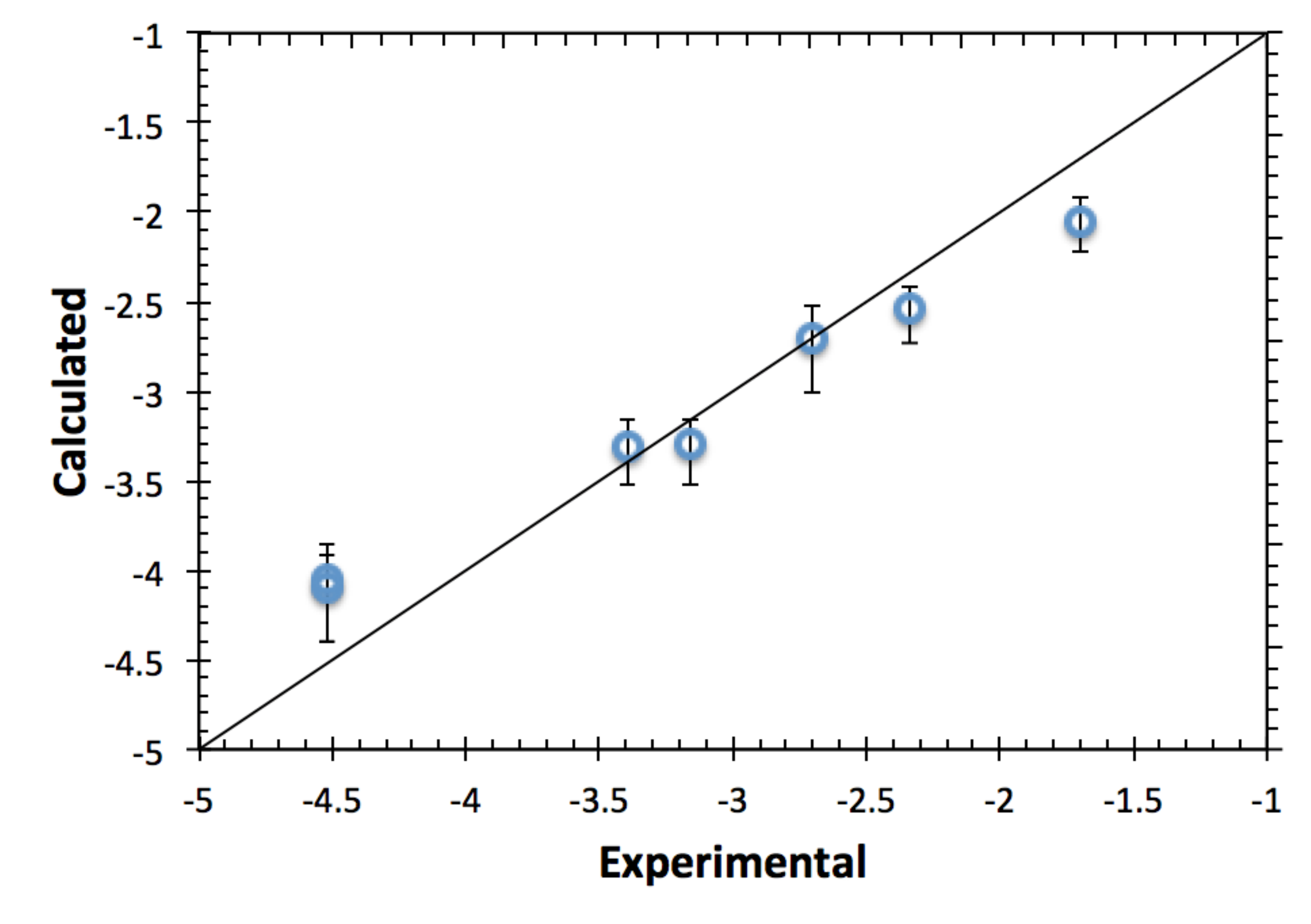}
\end{centering}
\caption{Experimental versus calculated critical micelle
  concentrations, CMC, (logarithmic scale, base 10) for selected alkyl
  ethoxylate surfactants (\CnEm\ family). Top right corresponds to
  molecules with high CMC values, bottom left to low CMC
  values. \label{fig:cmc}}
\end{figure}

We have also determined the limitations of the brute-force calculation
of $\logP$, and as such one can envisage extending the approach to
incorporate methods inspired by the molecular dynamics studies
discussed in the introduction.  In particular Monte Carlo methods like
Widom insertion could be employed to calculate the transfer free
energies, and these are likely to be successful, even for short
molecules, for the soft interactions that underpin the DPD model
\cite{wijmans_smit_groot_2001}.  Also, Gibbs ensemble methods could be
employed to generate equilibrated water and octanol phases eliminating
the need to control for the presence of an interface.  Perhaps most
intriguingly, standard Ornstein-Zernike integral equation closures
appear to be very accurate for DPD, and perhaps methods developed for
molecular liquids such as RISM or SAFT could be applied to calculate
phase equilibria and solute partitioning directly for the
coarse-grained model \cite{RISM-REVIEW, SAFT-orig}, provided the
appropriate wet octanol model is used to compare with experimental
data (correct experimental concentration of water in octanol).

We should also remark that whilst the present approach is presented as
a case study in DPD parametrization, it offers a route to calculate
\denovo\ $\logP$ values.  As such it may not be competitive in terms
of speed compared to the rapid empirical methods listed in the
introduction, but it does appear to be competitive in terms of
accuracy \cite{fraaije_2016}.

\section{Acknowledgements}\label{sec:ack}
The authors would like to thank the STFC Hartree Centre for supporting
this effort by dedicating human resource and computational
infrastructure to this work. RLA and DJB would like to pay particular
thanks to Ed Pyzer Knapp (IBM) and Annalaura Del Regno (STFC) for many
stimulating conversations that helped in progressing this work. This
work has been partially funded by Innovate UK project 101712 and the
authors are grateful to the other members of the UK Computer Aided
Formulation (CAF) consortium project for their input and for
stimulating discussions.

\appendix

\section{Methods}\label{app:meth}
We performed DPD simulations using the \DLMESO\ simulation package
\cite{dlmeso}. Reduced units are used in which all DPD beads have unit
mass, the temperature $\kT=1$, and the base length $\rc=1$
(\ie\ cutoff for the water bead self interaction).  For all simulations a
DPD time step of 0.01 (in reduced units) was adopted and trajectory
data was collected every 10 DPD time units (1000 time steps).

For an
up to date perspective on the DPD methodology see the recent work of
Espa\~nol and Warren \cite{espanol_warren_2017}.

\subsection{Liquid phase densities}
In the second optimization step described in the main text, we
performed constant pressure (NPT) simulations using the Langevin
piston implementation of Jakobsen \cite{jakobsen_2005}. The pressure
was set to match that of pure water in the model, which corresponds to
the pressure in a pure DPD fluid at reduced density $\rho=3$ and
repulsion amplitude $A=25$.  This was determined separately to be
$P=23.7\pm0.1$ (in DPD units).

The NPT simulations were carried out for molecules containing alkane,
alcohol, amine, ether and benzene moieties, and the self-interaction
parameters were varied as described in the main text to reproduce
experimental densities of the species investigated (where multiple
bead types were present in a test molecule, the arithmetic mixing rule
$A_{ij}=\frac{1}{2}(A_{ii}+A_{jj})$ was used as a first
approximation).  Simulations were run for 300 DPD time units and data
collected after the initial 150 DPD time units.

\subsection{Water-octanol partition coefficients}
In contrast to the simulations carried out to fit self-interactions to
experimental densities, for measuring $\logP$ we carried out constant
volume (NVT) simulations. As such initial simulation boxes were
constructed so that half of the volume of the simulation cell water
filled with water and the other half with octanol. Numbers of beads of
each type were selected in order to achieve the desired system-wide
target pressure of $P=23.7$. This is important to maintain the correct
density of the two solvent phases.

Four simulation box sizes were adopted in the calculation of $\logP$,
where each was a multiple of a basic $60\times20\times20$ box.  These
are listed in \Tabref{tab:boxes}.  The largest of these (`huge')
corresponds to $1.2\times10^6$ beads in total.  For the `small'
simulations, the left part of the box (the region $0<x<30$) was
populated by 36\,000 water beads and the right part ($30<x<60$) by
7800 octanol molecules (39\,000 beads). The total number of beads
therefore being 75\,000. For the larger boxes the small box was
replicated in the $y$ and $z$ dimensions as required. Therefore, the
simulation begins with water and octanol partitioned into their
respective pure phases. Solute molecules were added by random
insertion into the simulation box. The number of water molecules and
octanol molecules are correspondingly adjusted to account for the
presence of solute molecules. By following this approach all of our
simulations were performed within 2\% of the target pressure with the
exception of glycerol and tetritol which had pressures within 5\% of
the target value. The initial configurations were created using the
\PACKMOL\ package \cite{packmol}.

Different box sizes are required to cover different $\logP$ ranges. A
simple estimate suggests the maximum achievable $\logP$ range, for a
given simulation box size. For example, in a `small' box containing
5\% solute beads, there are a total of 3750 solute beads. Assuming a
solute molecule comprises of 3 beads, and supposing that of order one
molecule should be present in the \emph{disfavourable} phase at all times,
this corresponds to a maximum achievable range of $\logP=
\pm\log_{10}(3750/3)\approx \pm3$. However, we have found in practice
this overestimates the accessible $\logP$ range.  Our empirically
determined limits for a reliable $\logP$ calculation (with 5\% solute
beads) are given as the second column in \Tabref{tab:boxes}.  With
these limitations sampling errors can be kept at a sensible level
($<10$\% of the calculated mean solute concentration).

The effect of the solute concentration upon calculated $\logP$ quality
and equilibration time was also explored. Trial simulations of 1--5\%
solute were carried out. Given that the number of solute molecules
present in the box defines the maximum limit for $\logP$ (as discussed
above) it is preferential to have a large as possible value for the
solute concentration in the simulation. In addition, with larger
solute concentrations, shorter simulations can be carried
out. However, a too high value may adversely effect the integrity of
the bulk phases (\eg\ by spontaneous phase separation of the solute)
making any calculated $\logP$ worthless.

\begin{figure} 
\center
\includegraphics[width=3.6in]{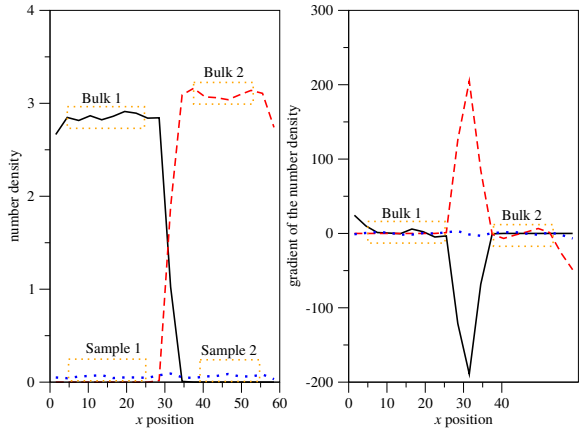}
\caption{Evaluating the region of interface and two bulk phases. The
  left subplot shows 1D number density profile for water (solid line)
  and octanol (dashed-line). The right subplot shows the gradient in
  number density for the equivalent profiles. A bulk phase is defined
  where the gradient varies by no more the $\pm \epsilon$ over more
  than one sample point (where $\epsilon$ = 0.05), with two such
  highlighted here by the dotted boxes. Regions that lie outside this
  are assumed to represent interfaces.\label{fig:interfaceplot}}
\end{figure}

A number of different options were trialled for initial solute
molecule placement: completely random (with $50:50$ mix in both
solvents), all in octanol, or all in water. In all cases particles
were randomly spread within the specified region. Positioning solute
molecules close to the interface was also trialled. We have found that
the ideal initial positioning of the solute depends upon the target
$\logP$. Whilst unsurprising, there are important consequences to
this. If, for example, glycerol is placed in any reasonably large
concentration in the octanol phase (supposing that the solute is
randomly distributed), it will spontaneously phase separate in this
phase. This has a big impact on the calculated $\logP$ value because
the system does not equilibrate over a feasible time scale. In our
trials for example we found the glycerol $\logP$ value fluctuates
wildly if the solute molecules are not initially positioned in the
bulk water region. Obviously if a simulation is allowed to run long
enough then the problem resolves itself, however, this is not
efficient for high throughput simulations for parametrization.

Throughout the course of our study, it was found that simulations only
needed to be run for a maximum of $20\,000$ DPD time units to meet the
criteria for equilibration and data collection of $\logP$ values for
the `small' and `medium' box sizes. Larger boxes are used for higher
$\logP$ values and as such the actual number of steps to reach
equilibrium was larger and typically took around 30\,000 DPD time
units.  Indicative run times for the typical $\logP$ calculations are
given in the final columns of \Tabref{tab:boxes}.

\begin{table}
\caption{Simulation box sizes used in $\logP$ calculations.  The
  second column indicates the range of $\logP$ values accessible to
  brute-force simulation. The indicative run times correspond to the
  given resource allocation (cores) on an IBM NextScale system
  ($2\times12$ core Intel Xeon processors; Ivy Bridge E5-2697v2
  2.7\,GHz; 64\,GB RAM).\label{tab:boxes}}
\setlength\tabcolsep{0.5em}
\vspace{1ex}
\begin{tabularx}{0.48\textwidth}{lcccccc}
\hline\\[-6pt]
 \multicolumn{1}{c}{box size} && $\logP$ && run time && cores \\[3pt]
\hline\\[-6pt]
$60\times20\times20$ (small)  && $\pm2.0$ && 2\,hrs && \x96 \\
$60\times40\times40$ (medium) && $\pm2.5$ && 3\,hrs && 144 \\
$60\times60\times60$ (large)  && $\pm3.5$ && 5\,hrs && 240 \\
$60\times80\times80$ (huge)   && $\pm4.0$ && 6\,hrs && 512 \\[3pt]
\hline
\end{tabularx}
\end{table}

We turn now to the methodology of extracting $\logP$ values from
simulation data.  In this work we calculate the value of $\logP$ for a
molecule by direct application of \Eqref{eq:logP}, using computed
values of the solute concentrations in octanol and water phases.  For
each case we undertook the following procedure:
\begin{enumerate}   
  \item Calculate the 1D number density profile running along the
    $x$-axis (\ie\ normal to the interface), see left pane of 
    \Figref{fig:interfaceplot}.
  \item Determine where the interface lies along $x$.
  \item Determine the bulk solvent boundary $x$-positions and
    corresponding length of the bulk phase of water and octanol
    excluding the transitioning interface. These regions should be
    mutually exclusive.
  \item Calculate the average number of solute molecules within these
    two bulk regions. These provide our estimates of the mean number
    density of solute within the water and octanol phases.
  \item Calculate the ratio of these number densities to obtain
    $\logP$ from \Eqref{eq:logP}.
\end{enumerate}

In calculating the mean solute concentrations the interface region
between octanol and water was excluded to ensure only the bulk phase
concentrations should be considered, analogous to experiment. To
facilitate this the following algorithm was developed to automatically
identify the interface region. We first take the time-averaged 1D
number density profiles of water and of octanol and calculate the
gradient of these (see right pane of 
\Figref{fig:interfaceplot}). The regions where the gradient fluctuates
around zero define the bulk phases. The interface region can be
identified by looking for a spike (positive or negative) in the
gradient that is an order of magnitude greater than the fluctuations
seen in the bulk regions. This spike defines the interface region to
be excluded from number density calculations. The left-hand pane of
\Figref{fig:interfaceplot} shows how the number density of the two
bulk solvents varies along the $x$-axis. The regions over which the
solute concentration value can be calculated is highlighted for each
bulk solvent. Estimates for the number density of solute within bulk
phase were calculated by taking the mean value within the boundaries
of each bulk region (labelled Sample 1 and Sample 2 in 
\Figref{fig:interfaceplot}).

To estimate the error in $\logP$ calculated in this way, we use block
averaging of the time series of $\logP$ obtained from the
instantaneous mean solute concentrations from the recorded trajectory
(\ie\ every 10 DPD time units).  The block size is 100 DPD time units
(10 measurements), and we average over 100 blocks to obtain a sample
mean value for the block.  The reported errors represent the standard
deviation in the calculated sample means over the whole data
collection period.

There are three important considerations in tuning our methods for
calculating $\logP$: handling poor parameter choice, poor
equilibration, and poor sampling of solute concentrations.  We discuss
each of these in turn.

\textit{Handling poor parameter choice} --- During the tuning of the
model interaction parameters to fit the $\logP$ of our test set there
was the potential for the bulk water-octanol phases to break down due
to poorly chosen model parameters. If the parameters were good then
there would be only two bulk phases (corresponding to how we
constructed the initial systems). If the parameters were poor then
either a single phase would be found, which suggests that water and
octanol had completely mixed, or more than two bulk phases would be
identified which suggested that some kind of microphase separation
might be happening.  Only when two distinct bulk phases are present
can the $\logP$ calculation be estimated with confidence using our
methodology.

\textit{Poor equilibration} --- Simulations were started with
molecular configurations that may be far from equilibrium with respect
to $\logP$, such as uniformly dispersed solute molecules for
example. Over time, the solute molecules migrate to their preferred
phase and during that period the estimated value of $\logP$ value will
change. Obviously for obtaining thermodynamic averages, data should
not be used until the system has equilibrated.  In this work systems
were considered equilibrated after the estimated value of $\logP$
remained stable over 100 blocks of time (where as above each block is
100 DPD time units).  The choice of 100 blocks was determined somewhat
arbitrarily, but seems to represents a good choice for the systems and
system sizes considered in this work. Data was collected for 10000 DPD
time units following equilibration (with particle data collected once
every 10 DPD time units) in order to determine the average (sample
mean) $\logP$ value for a particular system. If the standard deviation
of a particular $\logP$ was $>10$\% of the mean value we considered
the value to be void and sampled the system in a larger simulation box
to reduce error. We discuss typical equilibration times in the results
section for each of the simulation box sizes considered. As discussed
in the previous section, when the simulation box is comparably small
relative to the true $\logP$, all the solute could tend to accumulate in one
of the solvents. In these cases we rejected the results of the
calculations and re-ran the simulation in a larger simulation box.

\textit{Poor sampling} --- To calculate $\logP$ reliably from our
simulations it is important to ensure good sampling, and therefore get
good estimates, of the mean solute concentrations.  This is
essentially a problem of counting statistics, and critically depends
on the number of solute molecules in the \emph{disfavourable} phase.
It provides practical limits on the overall solute concentration, and
on the $\logP$ range that can be measured for a given simulation box
size.  Pragmatically we found that reliable estimates can be obtained
for $\logP$ using the above block averaging scheme (over 100 blocks of
10 DPD time units), using 5\% solute concentration, provided that we
specify if the standard deviation is greater than 10\% of the sample
mean of the $\logP$ we reject the measurement and re-do the
calculation in the next larger simulation box. A 5\% solute
concentration (\ie\ 5\% of beads in the box comprise the solute
molecules) equates to a mole fraction of 0.025 for the smallest
molecule we consider (2 beads) and 0.01 for the largest (5
beads). Should large molecules be sampled, which comprise a large
number of beads, there may be insufficient solute material to achieve
good sampling. We recommend that mole fraction values be in the range
of that specified above for further fitting.

\subsection{Critical micelle concentrations}
Constant pressure simulations were performed to calculate the CMC of
seven non-ionic surfactants of the \CnEm\ family. Specifically,
$\mathrm{C_{6}E_{4}}$, $\mathrm{C_{8}E_{4}}$, $\mathrm{C_{8}E_{8}}$,
$\mathrm{C_{10}E_{6}}$, $\mathrm{C_{10}E_{9}}$, $\mathrm{C_{12}E_{6}}$
and $\mathrm{C_{12}E_{7}}$ surfactants were explored. Simulation boxes
contained 325\,000 DPD beads and were run for 30\,000 DPD time
units. The initial $1/3$ of simulation time was used for equilibration
and the subsequent $2/3$ for data collection. The CMC was extracted
adopting the method outlined in Johnston \etal\ \cite{cmchartree}.

\section{Note on compressibility matching}\label{app:comp}
The baseline choice $A=25$ for the water bead repulsion amplitude
originated in the seminal work of Groot and Warren \cite{grootwarren},
who attempted to match the compressibility of DPD water to that of
`real' water.  It was later found that there was a missing factor of
the mapping number $\Nmap$, so that for a correct compressibility
matching one should use $A\approx 26\,\Nmap$ (see below)
\cite{grootrabone_2001}. Since our preference is not to do this, a few
words of explanation are warranted.  We first note that a large value
of the baseline repulsion leads to increased solvent structuring,
which can be regarded as a discretization artifact of the coarse
grained model.  Indeed, if $A\agt200$ the DPD solvent `freezes' (most
likely into a BCC structure) \cite{hartmut}.  Thus we should prefer to
use a small baseline repulsion amplitude; the question is: how small?

To answer this, recall that from statistical mechanics the relative
mean square density fluctuations in a volume $V$ are
$\langle\Delta\rho^2\rangle/\rho^2=\kt\kT/V$ where
$\kt^{-1}=\rho\,\partial p/\partial\rho$ is the (inverse) isothermal
compressibility.  Thus relative density fluctuations are inversely
proportional to the sample volume, and for liquid water at room
temperature the coefficient of proportionality $\kt\kT\approx
1.9\,\angscube$.  For DPD the same quantity is also well defined.
Assuming the equation of state $p\approx\alpha A\rho^2$ with
$\alpha\approx0.101$ \cite{grootwarren} (\ie\ neglecting the ideal
contribution which is small), one has $\kt\kT\approx 1/(2\alpha
A\rho^2)$, in DPD units.  Recalling that $\rho\Nmap\vmol\equiv1$ one
finally has the formal mapping $\kt\kT=\Nmap\vmol/(2\alpha
A\rho\rc^3)$.  Making the standard choice $\rho\rc^3=3$ and inserting
the actual numbers for water arrives at $A\approx 26\,\Nmap$ as
claimed.

The argument could obviously be extended to match compressibility, and
hence density fluctuations, for other molecular liquids.  However,
rather than insisting that the compressibility be exactly matched as
in earlier works, we here argue what really matters is that density
fluctuations should be relatively insignificant \emph{above the DPD
  length scale}.  For example in water
$\langle\Delta\rho^2\rangle/\rho^2\alt0.01$ for $V\agt\rc^3\approx
180\,\angscube$.  To ensure that the relative density fluctuations in
the DPD water model are bounded by
$\langle\Delta\rho^2\rangle/\rho^2\alt0.05$ (for example), requires
only that $1/2\alpha A\rho^2\alt0.05$, or $A\agt 10$ for standard DPD.
This is satisfied for our purposes for all our bead types.

%\bibliographystyle{aipnum4-1}
%\bibliography{references.bib}

\end{document}